\documentclass[aps,prl,showpacs,floats,twocolumn,amsfonts,amssymb,unsortedaddress,floatfix]{revtex4}
\usepackage{graphicx}
\usepackage{amssymb}
\usepackage{bm}

\input epsf
\usepackage{bm}
\usepackage{amsmath}
\usepackage{subfigure}

\begin{document}

\renewcommand{\bottomfraction}{0.7}
\renewcommand{\topfraction}{0.7}
\renewcommand{\textfraction}{0.2}
\renewcommand{\floatpagefraction}{0.7}
\renewcommand{\thesection}{\arabic{section}}
\addtolength{\topmargin}{10pt}
\def\Bbb{\mathbb}
\def\thinspace{\kern .16667em}
\def\punto{\thinspace .\thinspace}

\title{Static and Dynamical Phyllotaxis in a Magnetic Cactus.}
\author{Cristiano Nisoli$^{1,3}$, Nathaniel M. Gabor$^{2,3}$, Paul E. Lammert$^{3}$, J. D. Maynard$^{3}$ and Vincent H. Crespi$^{3}$.}

\affiliation{$^{1}$ CNLS and T-Division,  Los Alamos National Laboratory, Los Alamos NM 87545 \\ 
$^{2}$Department of Physics 
Cornell University, 109 Clark Hall, Ithaca, NY 14853-2501 \\ 
$^{3}$Department of Physics 
The Pennsylvania State University, University Park, PA 16802-6300}
\date{\today}

\begin{abstract}
While the statics of many simple physical systems reproduce the striking number-theoretical patterns found in the phyllotaxis of living beings, their dynamics reveal unusual excitations: multiple classical rotons and a large family of interconverting topological solitons. As we introduce those, we also demonstrate experimentally for the first time Levitov's celebrated model for phyllotaxis. Applications at different scales and in different areas of physics are proposed and discussed.
\end{abstract}
\maketitle

{\it Phyllotaxis}, the study of mathematical regularities in plants, challenged Kepler and da Vinci~\cite{Grew}, inspired the Bravais lattice of crystallography~\cite{Bravais}, and may well have motivated humanity's first mathematical inquiries~\cite{Grew}. In groundbreaking work, Levitov proposed~\cite{Levitov} that the appearance of the Fibonacci sequence and golden mean in the disposition of spines on a cactus is replicated in physics, in the {\it statics} of cylindrically constrained, repulsive objects~\cite{Levitov}. Here we prove experimentally Levitov's model, and describe for the first time the intriguing collective excitations of the phyllotactic geometry: multiple classical rotons and a huge family of interconverting topological solitons.  A simple geometrical mismatch underpins all these phenomena: nearest neighbors in one dimension (i.e. along the axis) are not nearest neighbors in the full three dimensions. Being purely geometrical in origin, dynamical phyllotaxis could occur in many physical systems, including trapped ions in cylindrical potentials~\cite{LoeschScheel}, Wigner crystals in curved nanostructures, or crystalized ion beams~\cite{Schiffer, Pallas}.

\begin{figure}[t!]
\begin{center}
\includegraphics[width=2.9 in]{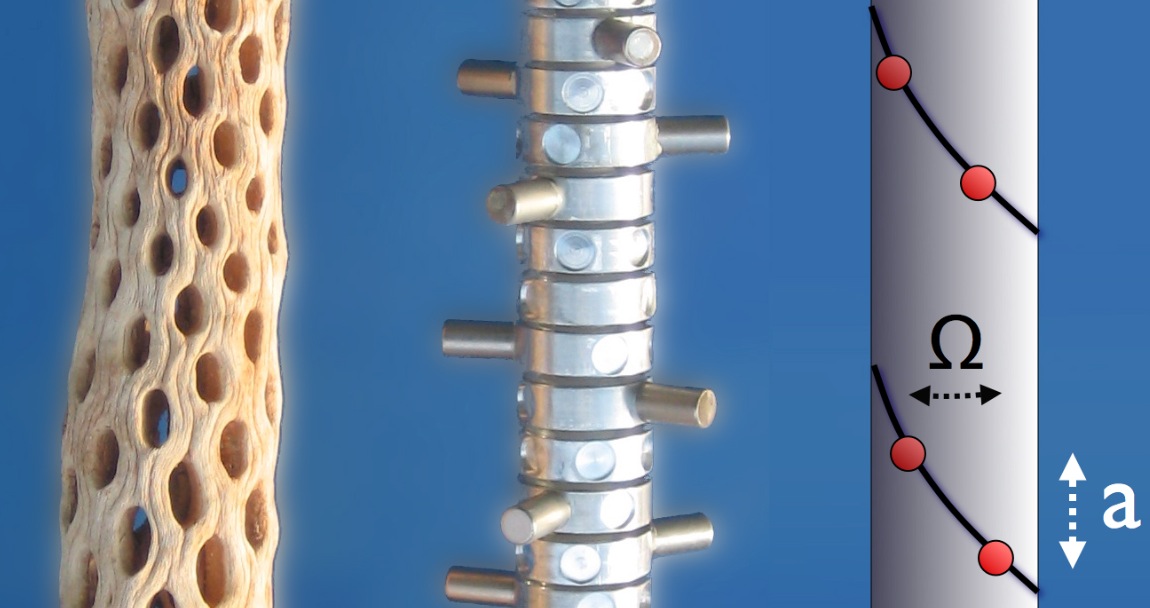}\vspace{-4mm}
\end{center}
\caption{
A specimen of {\it Mammillaria elongata} displaying a helical morphology ubiquitous to nature, a magnetic cactus of dipoles on stacked bearings, and a schematic of a wrapped Bravais lattice showing the angular offset (screw angle) $\Omega$ and the axial separation $a$ between particles.}\vspace{-4mm}
\label{structure}
\end{figure}

First we review {\it static} phyllotaxis as in Levitov~\cite{Levitov}: assume that a set of particles with long range repulsive interactions, when confined to a cylindrical shell of radius $R$, forms a helix with a fixed angular offset $\Omega$ between particles and a uniform axial spacing $a$, as in Figure~\ref{structure}. At low ($a/R \gg 1$) linear particle density, the optimal angular offsets that  maximizes distance between neighboring particles is $\Omega = \pi$. However, when $a/R$ drops below $2/\sqrt{3}$, the {\it second} neighbor in the axial coordinate (or ``axial index'') becomes the {\it first} neighbor in three-dimensional space, and so $\Omega = \pi$ becomes unfavorable. As density increases further, the angles $2\pi/3$ and $4\pi/3$ also become unfavorable due to third-neighbor interactions and so on, eventually generating a so-called Farey tree of unfavorable angles~\cite{Farey}, where every new fractional multiple of $2\pi$ is made by summing the numerators and denominators of the two previously adjacent multiples~\cite{Levitov}. 
A growing plant evolves quasi-statically from one optimal $\Omega$ to another as $R/a$ increases, asymptoting to the golden angle  $\Omega_1=2\pi/\left(\tau +1\right)$ $\left[\tau=\left(1+\sqrt{5}\right)/2\right]$, ubiquitous in botany. Occasional ``wrong turns'' at bifurcations during growth lead to alternative angles, the most common of which, $\Omega_p=2\pi/\left(\tau +p\right)$ with $p= 2, 3$, are called in botany second and third phyllotaxis.

\begin{figure}[t!]
\begin{center}
\vspace{-3mm}\includegraphics[width=3. in]{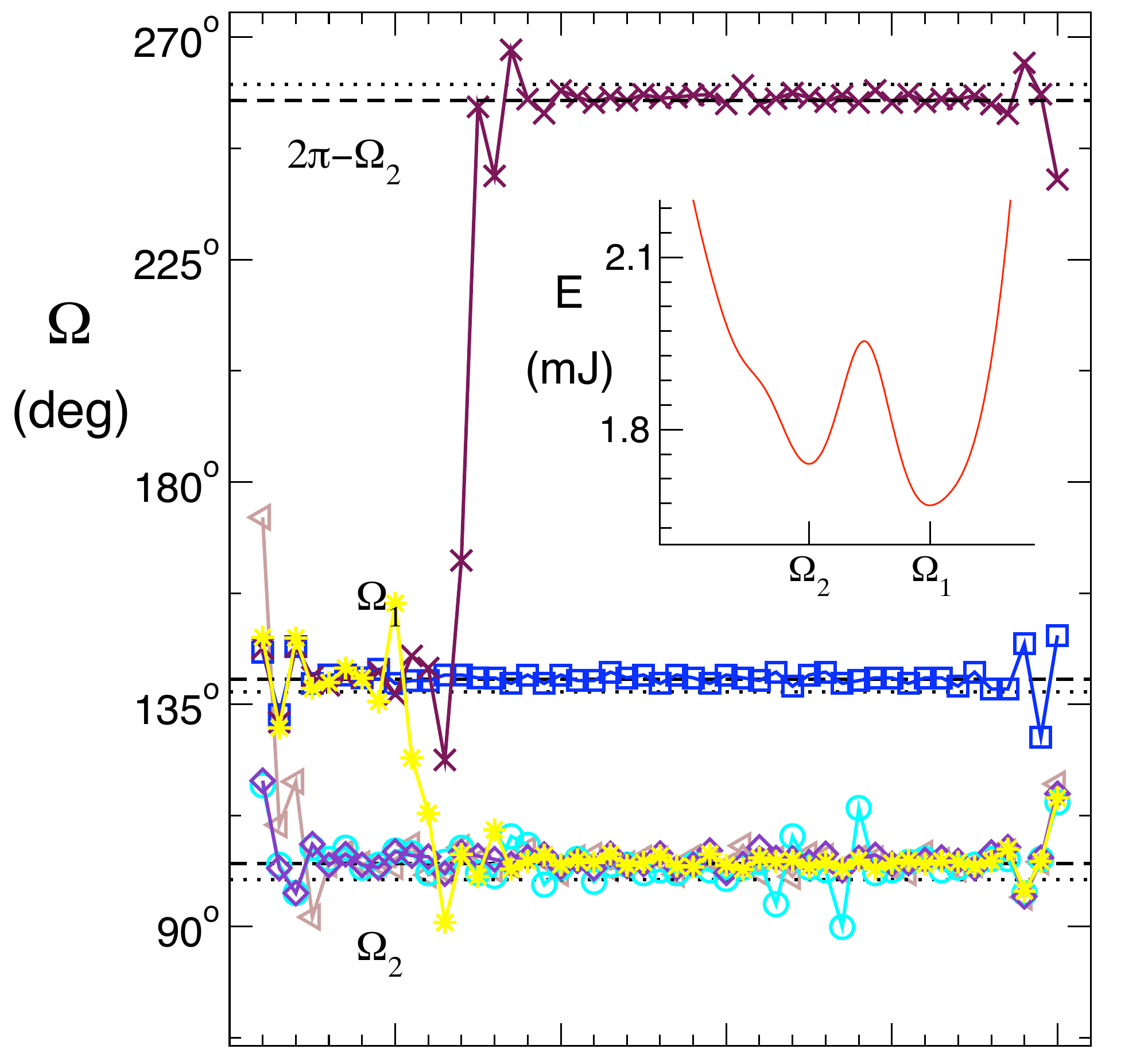}\vspace{-.5 mm}
\includegraphics[width=3. in]{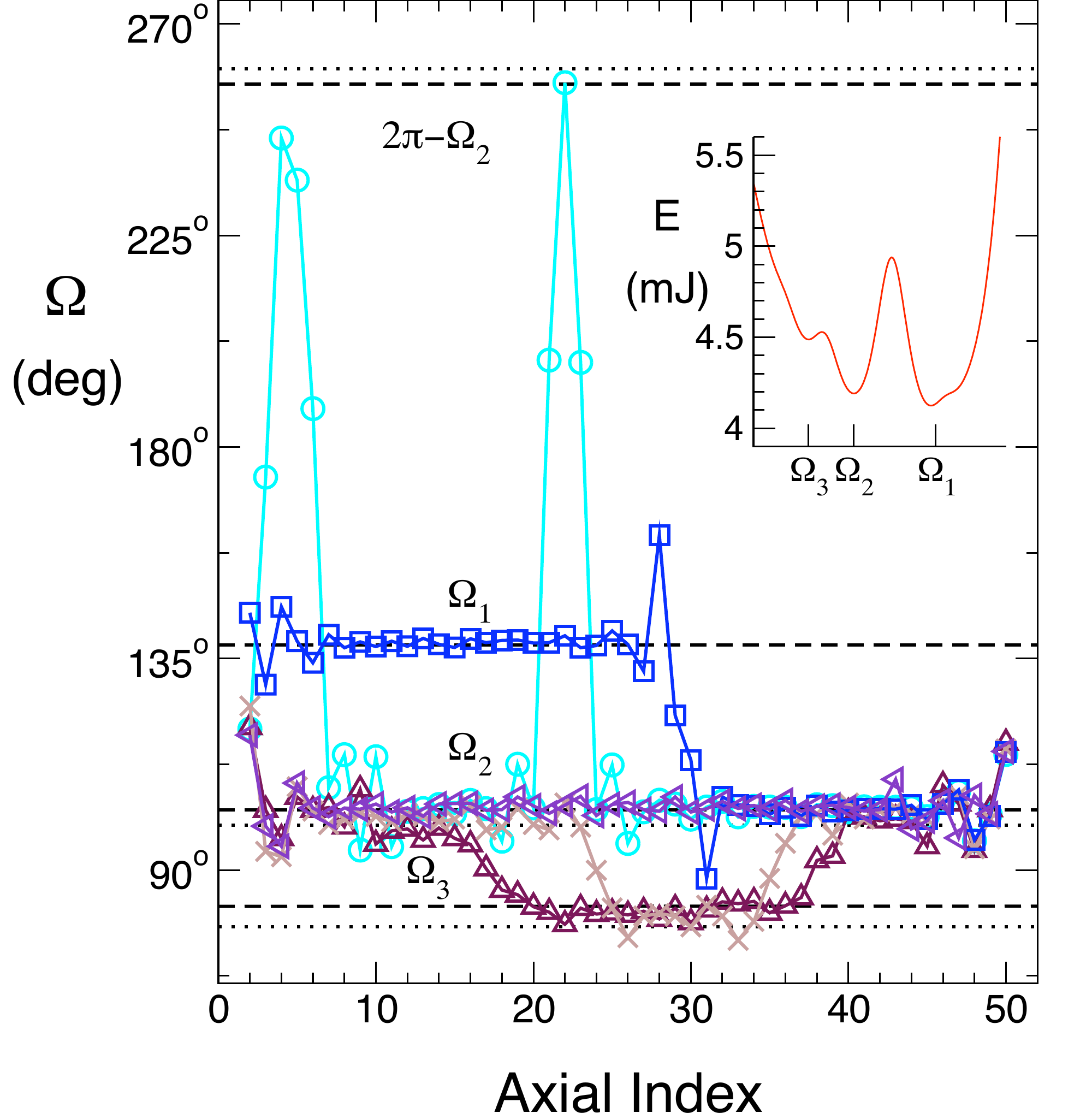}\vspace{-3mm} 
\caption{
Measured angular offsets between successive magnets for magnetic cacti with short (top) and long (bottom) magnets, plotted versus axial index. Flat regions are perfect spirals and steps are kinks between domains of different spiralling angles. Dotted lines give the phyllotactic angles $\Omega_1$, $\Omega_2$, $\Omega_3$ and $2\pi-\Omega_2$ defined in the text. Dashed lines are minima of the magnetic lattice energy (insets) calculated by interpolating the measured pair-wise magnet-magnet interaction. The magnetic cactus recapitulates botany.}\vspace{-5mm}
\end{center}
\label{expdata}
\end{figure}
\begin{figure}[t!!]
\center
\vspace{-5mm}\hspace{1mm}\includegraphics[width=3. in]{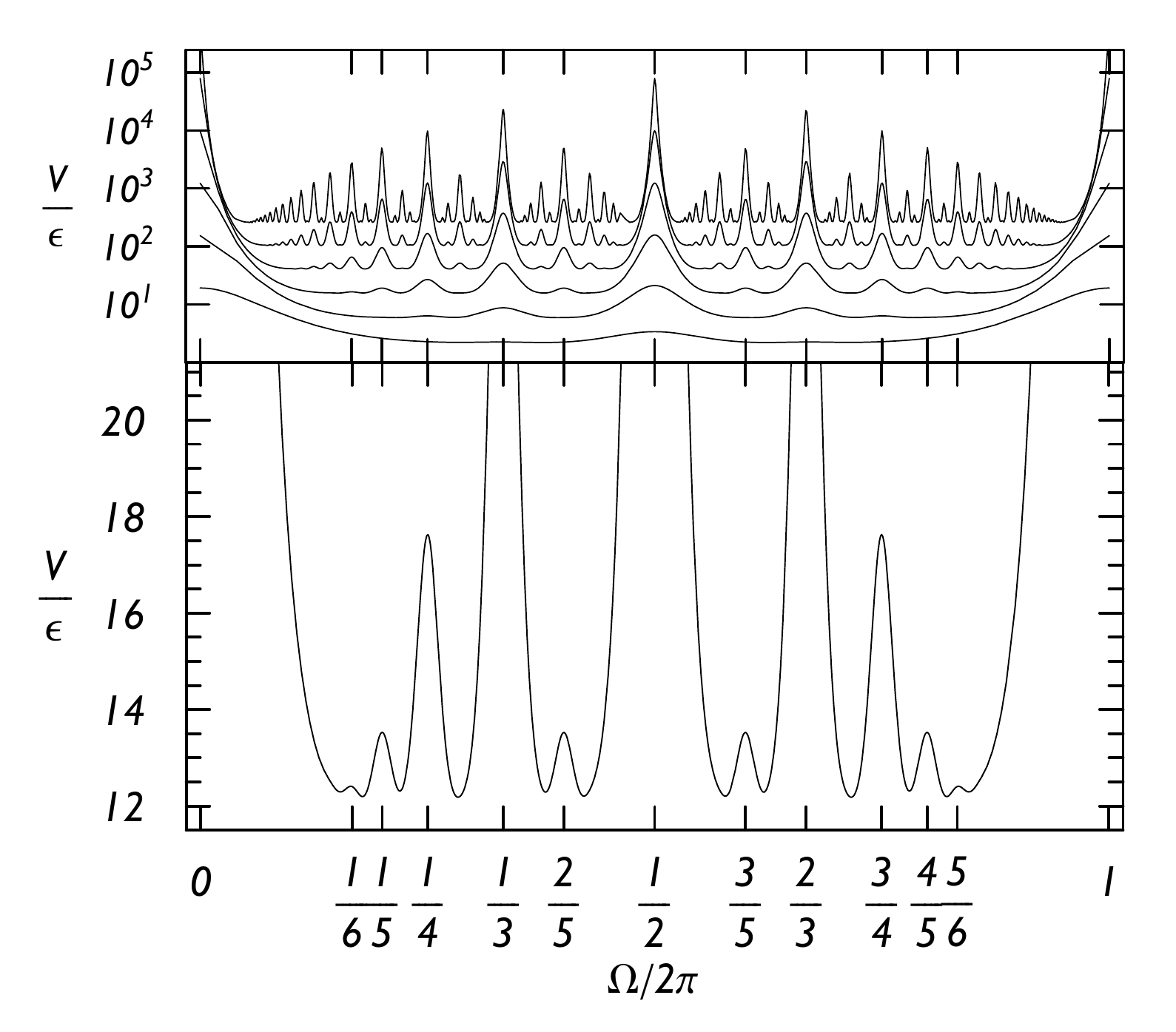}\vspace{-4mm}
\caption{Top: Lattice energy $V(\Omega)$ (in units of $\epsilon=p^2 R^{-3}$) versus screw angle for successively halving values of $a/R$ starting from $0.5$, showing more highly degenerate curves. Bottom: energy curve for $a/R=0.15$, with six distinct minima. The highest rank observed, $J=6$, matches Eq.~\ref{J}}\vspace{-5 mm}
\label{spectra}
\end{figure}

This appealing scenario requires experimental confirmation, since it rests on the assumption that the ground states are helical Bravais lattices parameterized by the angular offset $\Omega$.  Previous experimental reproductions of phyllotactic patterns did not interrogate the ground state of particles with long-ranged interactions; instead they examined kinetic processes in ferro-fluid droplets~\cite{Douady} or packings of hard spheres~\cite{Airy}. To properly test Levitov's model, we constructed two versions of a {\it magnetic cactus} consisting of 50 outward pointing, dipolar permanent magnets (spines) mounted on stacked coaxial bearings free to rotate about a vertical axis (stem), as in Figure~\ref{structure}. The system can be annealed into a lower energy state by mechanical agitation~\cite{EPAPS}.  We find that the final states so obtained consistently assume phyllotactic spirals whose angles $\Omega$ precisely reproduce first ($\Omega_1$), second ($\Omega_2$) and -- for a cactus with longer spines/dipoles -- third phyllotaxis ($\Omega_3$), as expected in a biological cactus and  as shown in Figure~2. Often, as also seen in Botany~\cite{kink},
the experimental system fragments into spirals of different $\Omega_p$, corresponding to two or more local energy minima separated by a domain wall; this is expected given the degeneracy and low dimensionality. 
Numerical optimization via a structural genetic algorithm confirms and extends these results in  axially unconstrained systems~\cite{EPAPS}.

Unlike its botanical analogue, the magnetic cactus can access {\it dynamics}. To better understand it, we first describe its fascinating self-similar energy landscape~\cite{Levitov, Farey}.  The essential results  apply equally to any smooth moderate-ranged repulsive interaction. We consider dipoles ${\bm p}$ directed radially outward, interacting via $v_{i,j}={\bm p}_i \cdot {\bm p}_j/ r_{i,j}^3 -3 ({\bm p}_i \cdot {\bm r_{i,j}})({\bm p}_j\cdot {\bm r_{i,j}}) /r_{i,j}^5$. The dipoles can rotate on their angular coordinate $\theta_n$ but are constrained axially to equal spacings $z_n=a n$, $n$ the axial index. $V= 1/2 \sum_{n \neq m} v_{n,m}$ is the total potential energy. To study dynamics,  kinetic energy can be added: $E=\frac{1}{2} I \sum_i {\dot {\theta}}_{i}^{~ 2} + V$. Figure~\ref{spectra} plots the lattice energy $V(\Omega)$ versus  angular offset  $\Omega$  for spiral lattices (i.e. $\theta_{n}=\Omega \!\ n$) at various densities. As discussed above, the lattice energy develops nearly degenerate peaks corresponding to commensurate spirals $\Omega=2\pi~i/j$ with $i, j$ coprime (i.e. having no common divisors): for these structures $\theta_k=\theta_{k+j}$, so $V$ is dominated by particles facing each other at a distance $j a$; we call these peaks of rank $j$. For  given $a/R$, there is a maximum rank $J$~\cite{EPAPS} (corresponding to the smallest peaks in Fig.~\ref{spectra}), 
\begin{equation}
J = \Big\lbrack{\kern -0.1 em}\Big\lbrack 
\sqrt{\frac{2\pi R}{a}} \Big\rbrack{\kern -0.12 em}\Big\rbrack
\label{J}
\end{equation}
$ \lbrack{\kern -0.1 em}\lbrack ~\rbrack{\kern -0.12 em}\rbrack$ denoting the integer part. The minima between the peaks also become nearly degenerate with increasing density, as each particle sees the others incommensurately averaged.  Via number-theoretical considerations the degeneracy is found  $D=\frac{3}{\pi^2} J^2 +O\left(J \log J\right)\propto \frac{6 R}{\pi a} $~\cite{EPAPS}.  A stable structure corresponding to a minimum bracketed by peaks of rank $j_1$ and $j_2$ is a spiraling lattice where nearest neighbors are at axial displacements $\pm a j_1$, $\pm a j_2$ and second nearest neighbors are at $\pm a (j_1+j_2)$ or $\pm a (j_1-j_2)$~\cite{Levitov}. Thus  $j_1$ and $j_2$, which are coprime~\cite{Farey},  give the number of crossing helices (in Botany, parastichies) that cover the lattice by connecting nearest neighbors. 
 
\begin{figure}[t!!]
\center
\vspace{-3mm}\includegraphics[width=3 in]{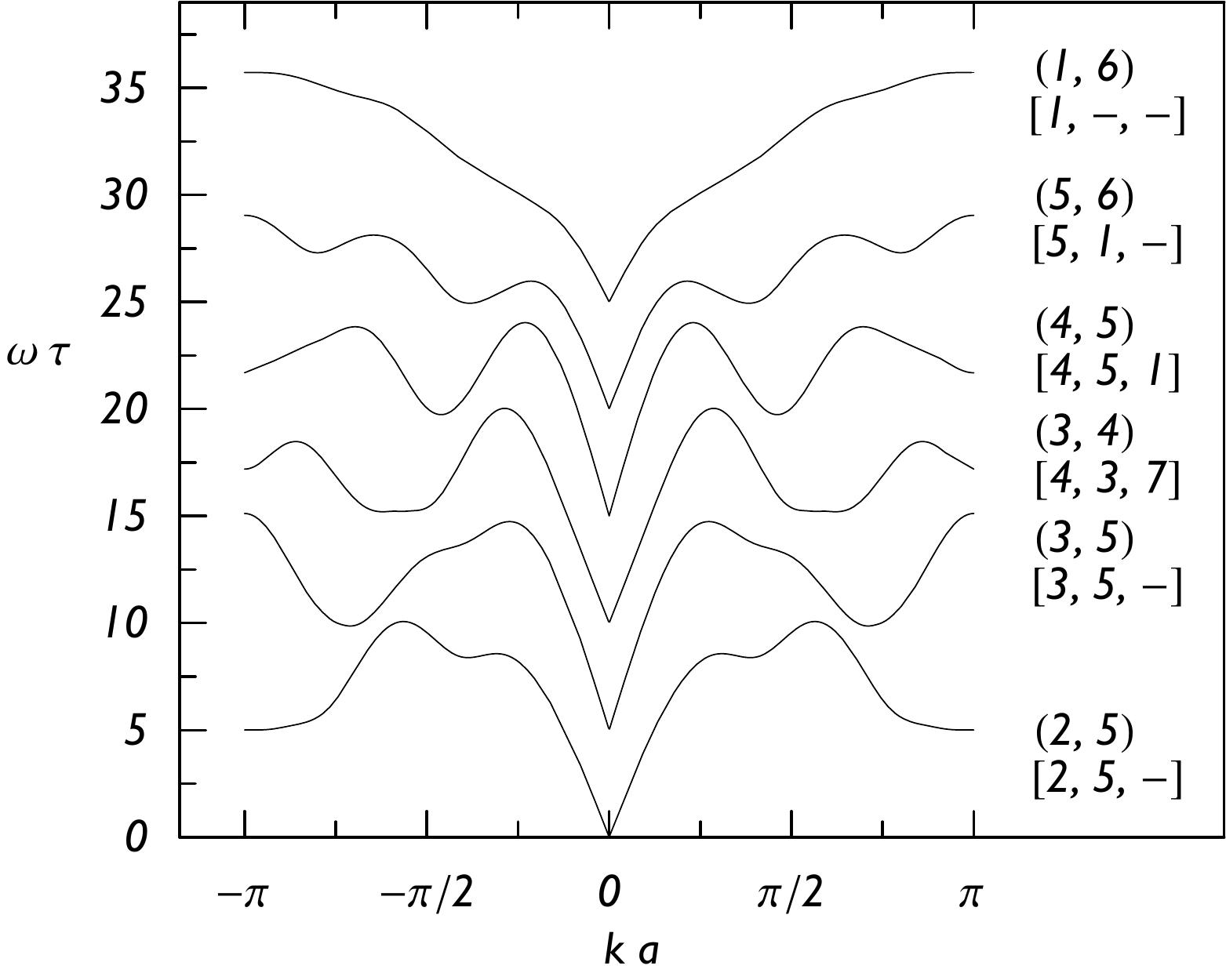}\vspace{-4mm}
\caption{
The phonon dispersion relations for the six phyllotactic lattices of Figure~\ref{spectra}, with multiple rotons and maxons. Corresponding parastichy numbers $(j_1,~j_2)$ (defined in the text) and the three most strongly interacting neighbors $[\tilde j_1,~\tilde j_2,~\tilde j_3]$ ($\omega_1\tilde j_1>\omega_2 \tilde j_2>\omega_3 \tilde j_3$) are given. The simple estimate for the number of rotons and maxons, $2 \tilde j_1 -1$, holds for all but (2,5). Each spectrum is offset by $5\tau^{-1}$ for clarity, where $\tau = \epsilon^{-1/2}I^{1/2}$ is the unit of time~\cite{EPAPS}.}\vspace{-4mm}
\label{phonons}
\end{figure}

This striking potential energy landscape 
returns non-monotonic dispersion relations in {\it linearized dynamics},  with many classical rotons and maxons. 
While rotons in superfluid $^4$He are associated with density fluctuations and generally held to be intrinsically quantum mechanical \cite{Roton}, phyllotactic rotons have a new, purely geometrical and classical origin.
 Consider oscillations around a stable spiral $ \Omega$: $\theta_n= \Omega \!\ n + \Psi_k \cos\left[k n-\omega(k) t\right]$. The normal mode frequency is
\begin{equation}
\omega(k)^2= 2\omega^2_{j_1}[1-\cos(j_1 k)] + 2\omega^2_{j_2}[1-\cos(j_2 k)] + \cdots
\label{dispersion}
\end{equation} 
where $j_m$ is the axial index of the $m$-th neighbor $\theta_{j_m}$  and $\omega_{j_m}$ is the second derivative of $v_{0,j_m}$ about the equilibrium spiral ${\Omega}$. For simplicity, let us truncate this expansion at nearest neighbors. Crucially,  nearest neighbors in the axial coordinate are {\it not} necessarily nearest neighbors in three dimensions, thus $\{j_1,j_2,j_3,\dots\}\ne\{1,2,3,\dots\}$. As explained before, the nearest neighbors $j_1,j_2$ are instead the parastichy numbers for $ \Omega$, and thus coprime~\cite{Farey,Levitov}: it follows that the dispersion relation of Eq.~\ref{dispersion}, truncated to just $(j_1, j_2)$, has a periodicity interval $[-\pi,\pi)$ (which is {\it not\/} a Brillouin zone, since our lattice is axially aperiodic), and is non-monotonic, but instead shows $2j_1 - 1$ minima and maxima -- or rotons and maxons (we order $j_1,j_2$ so that $\omega_{j_1} j_1 > \omega_{j_2} j_2$). Figure~\ref{phonons} shows the exact dispersion curves for each of the stable structures in the top panel of Figure~\ref{spectra}: extrema follow the $2 j_1 -1$ rule in all but one case. Physically, why are there rotons? Although increasing axial wave-vector drives particles nearby along the axis increasingly {\it out} of phase, it can drive particles that are neighbors in three dimensions more nearly {\it in} phase. This discrepancy between nearest neighbors in 1D and 3D 
drives rotonic behavior. Simulations 
confirm the localization of energy and momentum that is expected for rotons~\cite{EPAPS}.
\begin{figure}[!!t]
\center
\includegraphics[width=2.8 in]{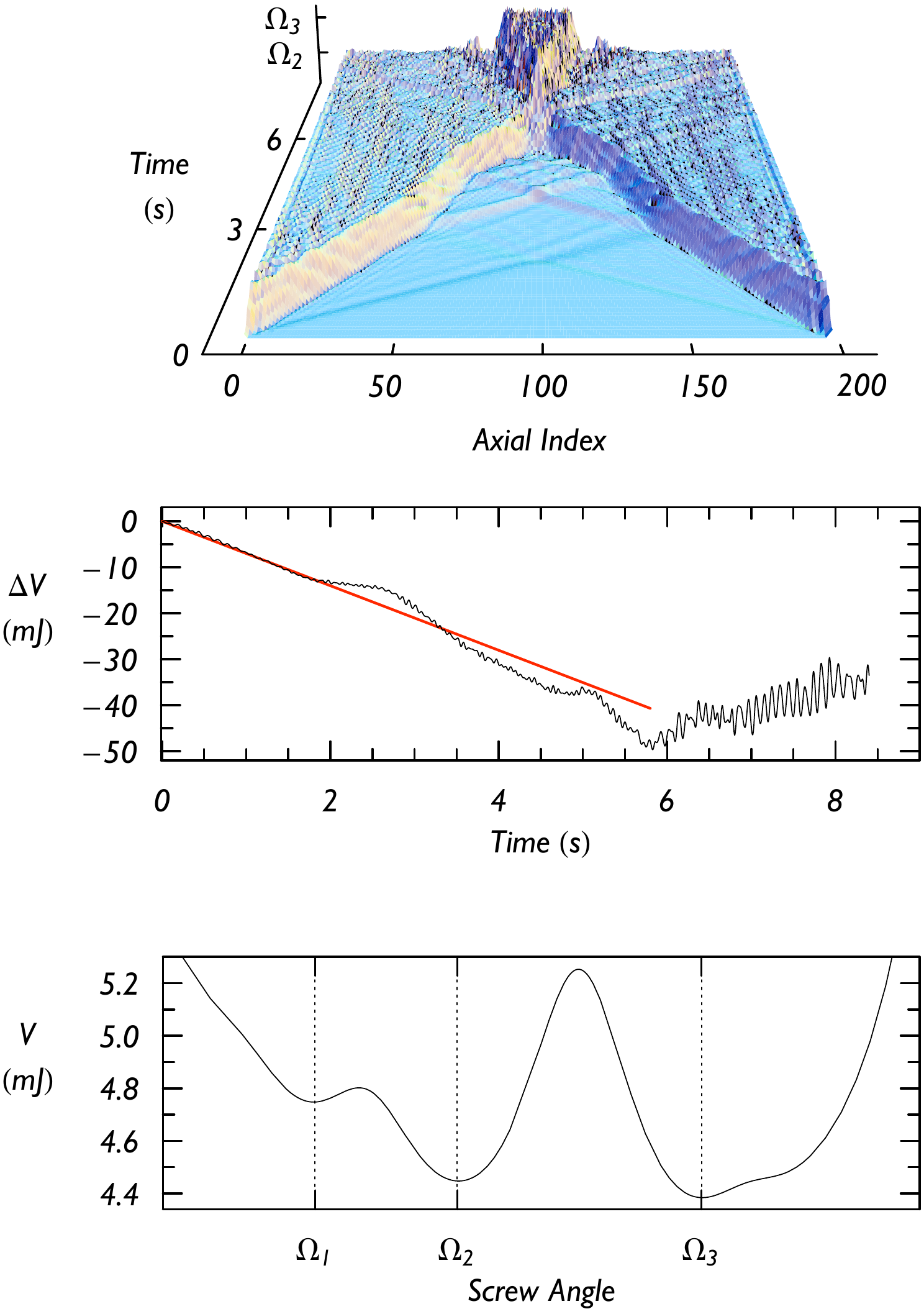}\vspace{-4mm}
\caption{
Two phyllotactic solitons approach, collide and transform. Top: The screw angle $\Omega$ plotted versus space and time for a collision of two phyllotactic solitons. Plateaux of different height correspond to spirals of different angle. Middle: The potential energy $V$ drops as the lower-energy domains advance, converting into rotational kinetic energy of the newborn domain (see online animation S5). Deviations starting around $t=2$ arise from precursor elastic waves propagating in front of the solitons that ensure local angular momentum conservation \protect\cite{NisoliPRE}. Bottom: The lattice energy $V(\Omega)$ corresponds to the physically realized  magnetic cactus of Figure~2 (bottom), with minima at $\Omega_1/2\pi=0.23$, $\Omega_2/2\pi=0.28$, $\Omega_3/2\pi=0.38$. For clarity, weak high-wave-vector phonons were smoothed by spatial averaging, here and in Figure~\protect\ref{protean}.}\vspace{-4mm}
\label{kinkrun}
\end{figure}

While the linear dynamics of phyllotaxis provides rotons, the {\it nonlinear} regime generates a new class of highly non-local topological {\it solitons} with unusually rich properties. One-dimensional degenerate systems are generally entropically unstable against domain wall formation~\cite{Lubensky}: we did see kinks both numerically and experimentally (Figure~2). Can these kinks travel as topological solitons? Experimentally, we observed the magnetic cactus expelling a higher energy domain by propagating its kink along the axis. Such axial motion of a kink between two domains of different helical angles confronts a dilemma: helical phase is unwound from one domain at a different rate than it is wound up by the other. Numerical simulations show that the moving domain wall solves this problem by placing adjacent domains into relative rotation (S5). If we compare this behaviour with the paradigmatic case of sine-Gordon-like 1-D topological solitons, which separate essentially equivalent  static domains and can trave at any subsonic speed~\cite{Lubensky}, we see that phyllotactic domain walls instead separate regions of different dynamics: energy and angular momentum {\it flow through} the topological soliton as it moves, rather than being concentrated in it, and its speed $v_K$ is tightly controlled by energy-momentum conservation, phase matching at the interface and boundary conditions. 

As a simple, symmetric example, consider a low-density system which reproduces our experimental cactus, prepared at rest in a metastable spiral at $\Omega_1$ with free boundaries. As depicted in Figure~\ref{kinkrun}, lower-energy domains of angle $\Omega_2$ spontaneously form at the edges and move inward  at a fixed speed while rotating uniformly at $\dot \theta=v_K \left(\Omega_2 - \Omega_1\right)$. As the rigid $\Omega_1-\Omega_2$ domain wall advances, it converts the potential energy difference $V_2-V_1$ into rotational kinetic energy: $I \left(\Omega_2 - \Omega_1\right)^2 v_k^3 t =2 \left(V_2-V_1\right) v_K t$, which requires a propagation speed of
\begin{equation}
v_{K}^2=\frac{2 \Delta V}{I \Delta \Omega^2}.
\label{speed}
\end{equation}
The measured  speed, $22.1 ~ \mathrm{a\  s^{-1}}$ (see Figure~\ref{kinkrun}), agrees well with the value predicted from Eq.~\ref{speed}, $23.4 ~ \mathrm{a \  s^{-1}}$, the discrepancy likely coming from phonon radiation (expected for a soliton on a discrete lattice).

\begin{figure}[!!t]
\center
\includegraphics[width=2.5 in]{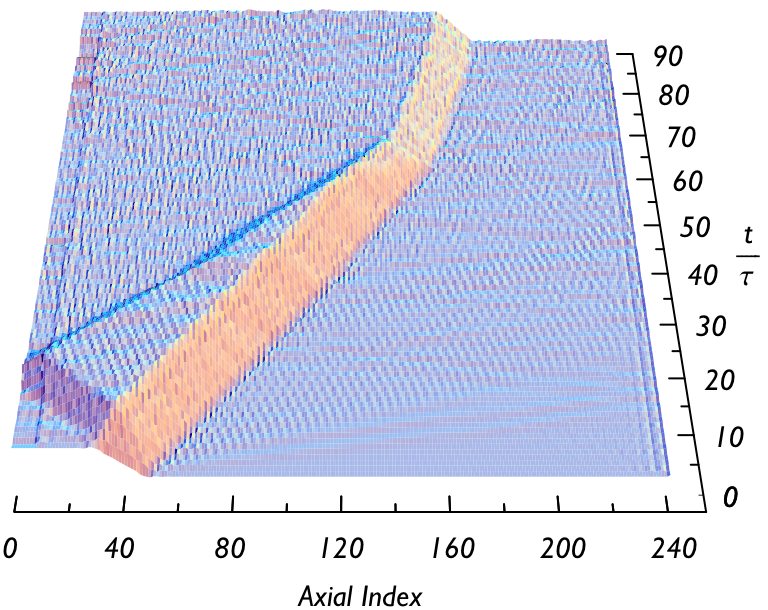}\vspace{-3mm}
\caption{The topological solitons can decay, merge, and change identity. An initial unstable domain boundary decays into two solitons. The left-hand soliton bounces off the free boundary, maintains its shape, propagates rightward faster than the other, collides with it, merges, and transforms into a third topological soliton with a different characteristic speed. S5 provides an animation.  The density is $a/R=0.15$.}\vspace{-4mm}
\label{protean}
\end{figure}

An extraordinarily rich phenomenology, well beyond that seen in traditional soliton models~\cite{Zabusky} emerges from more complex numerical simulations: kinks of different species merge, decay, change identity upon collision, and decompose at high temperature into a sea of constituent lattice particles. 
Figure~\ref{protean} depicts an unstable domain wall decaying into two topological solitons, one of which reflects off a free boundary then catches up and merges with the other to form a new soliton of different characteristic speed. A continuum analytical model that takes into account the rotational kinetic energy of the domains can predict and classify many of these phenomena~\cite{NisoliPRE}. 

Because dynamical phyllotaxis is purely geometrical in origin, this rich phenomenology of new excitations could appear across nearly every field of physics. Indeed phyllotactic domain walls have already been seen, but not recognized, in simulations~\cite{Schiffer} of cooled ion beams~\cite{Pallas} where the system self-organizes into concentric cylindrical shells. Trapped ions or dipolar molecules could attain a $J$  of several tens~\cite{LoeschScheel}. 
Colloidal particles on a cylindrical substrate provide a highly damped version~\cite{mesophase}, and polystyrene particles in air (as used to investigate~\cite{choi} the KTHNY theory of 2D melting~\cite{strandburg}) have reasonably low damping and long-range interaction. With charged repulsive particles and the constraint to equal axial spacing released, each topological soliton carries a different charge~\cite{NisoliPRE}. Strikingly, this charge is {\it not} conserved in soliton collisions, since it can be lent to or borrowed from a weakly pinned~\cite{Chitra} lattice. This could happen in Wigner crystals on suspended, weakly doped semiconducting carbon nanotubes or boron nitride nanotubes. Phyllotactic degeneracy requires small enough $a/R =r_s^2/2R^2$. Since $r_s\ge 37$  \cite{Tanatar}, from Eq.~\ref{J} we have that $R \ge\frac{37 J}{2 \sqrt{\pi}}  a_0 \simeq 5.5 J  ~ \mbox{\rm \AA}$ for a maximum peak rank $J$. $a_0$ is the Bohr radius for an appropriate effective mass and dielectric constant. If $a_0 = 0.53$ \AA, the minimal degeneracy, $D=2$ and $J=2$, occurs already for $R\simeq 11$ \AA. 
Finally, the extension of this fascinating geometrical structure to the quantum regime is particularly intriguing.

This work was supported by the National Science Foundation through DMR-0609243 and ECS-0609243.

\end{document}